# Enhancing sensitivity in absorption spectroscopy using a scattering cavity

JEONGHUN OH[1,2], KYEOREH LEE[1,2], AND YONGKEUN PARK[1,2,3,*]

[1]Department of Physics, Korea Advanced Institute of Science and Technology (KAIST), Daejeon, South Korea
[2]KAIST Institute for Health Science and Technology, KAIST, Daejeon, South Korea
[3]Tomocube Inc., Daejeon, South Korea
*Corresponding author: yk.park@kaist.ac.kr

**Absorption spectroscopy is widely used to detect samples with spectral specificity. Here, we propose and demonstrate a method for enhancing the sensitivity of absorption spectroscopy. Exploiting multiple light scattering generated by a boron nitride (h-BN) scattering cavity, the optical path lengths of light inside a diffusive reflective cavity are significantly increased, resulting in more than ten times enhancement of sensitivity in absorption spectroscopy. We demonstrate highly sensitive spectral measurements of low concentrations of malachite green and crystal violet aqueous solutions. Because this method only requires the addition of a scattering cavity to existing absorption spectroscopy, it is expected to enable immediate and widespread applications in various fields, from analytical chemistry to environmental sciences.**

Absorption spectroscopy detects a trace of a specific substance by measuring the absorbance of light at various wavelengths [1, 2]. Due to its instrumentational simplicity and high spectral specificity, absorption spectroscopy has been exploited in various fields, such as analytical chemistry [3, 4], atmospheric science [5, 6], geology [7], and atomic physics [8]. The measured absorbance is proportional to the optical path length (OPL) $l$ according to the Beer–Lambert law:

$$I = I_0 \exp(-A) = I_0 \exp(-\varepsilon c l), \quad \textbf{(1)}$$

where $I_0$ and $I$ are the measured intensities in the absence of a sample and in the presence of a sample, respectively, $A$ is the dimensionless absorbance of the sample, $\varepsilon$ is the molar attenuation coefficient of the sample, and $c$ is the molar concentration of the solution [9].

The sensitivity of absorption spectroscopy can be enhanced by increasing $l$. For instance, multi-path absorption cells with mirrors have been employed [10-13]. Such systems can significantly amplify the OPL but generally require a bulky system and sophisticated alignment. Attenuated total reflection spectroscopy exploits multiple total internal reflections in a crystal, which creates an evanescent wave [14, 15]. However, this method requires a complicated configuration, such as the exact optical contact between a sample and a crystal [16]. The need for a method of using simpler and more compact equipment to increase the OPL has been developed [17].

Multiple light scattering has been exploited to increase the OPL and overcome the limitations of conventional optical elements [18], including superlens [19], temperature [20], pressure sensing [21], wavelength meters [22], fiber-based spectrometers [23], and non-resonant lasers [24]. Recently, Martin *et al.* [25] suggested multiple scattering for enhancing the sensitivity of absorption spectroscopy. Dielectric microspheres were added to a sample solution to induce multiple scattering, which effectively increased the OPL. However, this method has various limitations: it requires the addition of microspheres for each measurement. Once mixed with microspheres for measurement, the remaining beads prevented the reuse of the sample. In addition, the degree of increase in the OPL varies depending on the quantity and size of the beads, as well as the refractive index (RI) difference between the microspheres and the sample. Thus, the enhancement factor varies according to the experimental environment.

Here, we present a simple but powerful method for increasing the OPL in absorption spectroscopy. By introducing a scattering cavity to an existing absorption spectroscopy to enclose a sample, the sensitivity of the measurements is significantly improved, because the light is trapped inside a scattering cavity and interacts with a sample numerous times before exiting. Unlike previous approaches, this method does not perturb a sample or require complex instruments. In addition, the method does not require calibration for each measurement or interfere with the sample condition.

In the conventional method, light passes through a cuvette only once. In the proposed method, the diffusive surface of a scattering cavity scatters light, causing the diffusively reflected light to pass through a longer path, and the reflected light is collected several times [Fig. 1(a)]. The experimental setup was composed of a halogen lamp (OSL1-EC, Thorlabs, Inc.), a custom-made scattering cavity, and a spectrometer (HR4000, Ocean Optics, Inc.) [Fig. 1(b)]. Two linear polarizers (LPVISE100-A, Thorlabs, Inc.) were employed as the beam power attenuators. A short-pass filter (under 750 nm, FES0750, Thorlabs, Inc.) was utilized to filter the wavelengths that exceeded the spectral range of the spectrometer. Note that linear polarizers and short-pass filters are not essential components.

The scattering cavity was made of >99.5% purity hexagonal boron nitride (h-BN). h-BN exhibits minimal absorbance [26, 27], and high diffuse reflectance in the visible region [28] make h-BN suitable for absorption spectroscopy. Due to its graphite-like atomic structure, h-BN exhibits excellent mechanical properties and exhibits satisfactory machinability [29-32]. The diffuse reflectance of h-BN was calibrated using a spectrophotometer (Lambda 1050, Perkin Elmer, Inc.) (Fig. 1(c)). h-BN has a high diffuse reflectance of more than 80% at wavelengths longer than 500 nm. The inverse adding-doubling method [33, 34] was used to estimate the absorption and scattering coefficients of h-BN. The estimated absorption $\mu_a$ and reduced scattering coefficients $\mu'_s$ were 0.023 mm$^{-1}$ and 129 mm$^{-1}$, respectively, at 532 nm. The front and top views of the scattering cavity are shown in Fig. 1(d). To prevent the direct passing of normal incident light, the exit hole was offset from the entrance with a height difference of 10 mm.

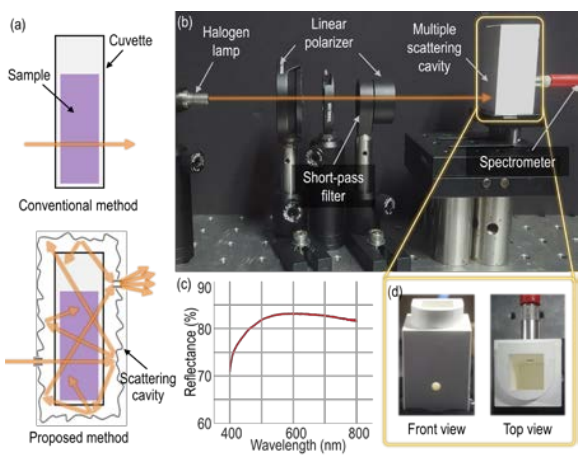

**Fig. 1.** (a) Schematics of the conventional method and proposed method. (b) Photograph of the experimental setup. (c) Reflectance of h-BN. (d) Front (left) view and top (right) view of the scattering cavity.

To verify this method, the absorption spectra of malachite green and crystal violet were measured. Malachite green and crystal violet have high water solubility and exhibit maximum absorption at wavelengths of 617 nm and 590 nm, respectively [35, 36]. We measured the malachite green aqueous solutions at various concentrations (Fig. 2). It was difficult to discern the malachite green solution with the naked eye when the concentration was below 1 μM. The absorption spectra were measured from the sample solution and deionized (DI) water, which correspond to $I$ and $I_0$, respectively, in Eq. (1). The normalized spectrum $I/I_o$ using the conventional method was measured in the same manner. Figure 2(b) presents the normalized spectra using the proposed and conventional methods for 0.2 μM malachite green aqueous solution. The absorption dip at 617 nm can be clearly observed in the proposed method, whereas the conventional method provides a low signal-to-noise ratio (SNR).

The normalized spectra of malachite green and crystal violet aqueous solutions at various concentrations are shown in Fig. 3. For validation, five measurements were acquired from the same sample solution (Figs. 3(a) and 3(e)) and five individual solutions of the same concentration (Figs. 3(c) and 3(g)). We discovered that the absorption dips were consistently observed at 617 nm and 590 nm for malachite green and crystal violet, respectively, which shows agreement with the known values. In addition, the linear relationship between concentration and absorbance in Eq. (1) is shown with high R-square values [Figs. 3(b), 3(d), 3(f), and 3(h)].

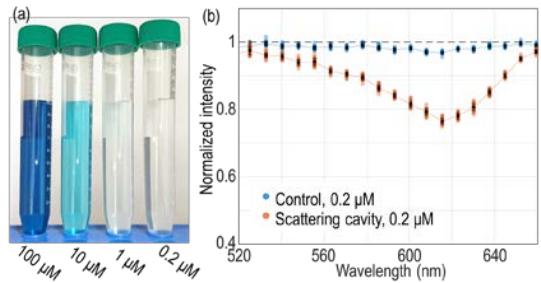

**Fig. 2.** (a) Photographs of malachite green aqueous solutions at 100, 10, 1.0, and 0.2 μM. (b) Comparison for the normalized absorption spectra of 0.2 μM malachite green aqueous solution measured with the conventional method and the proposed method. Error bars, standard deviations.

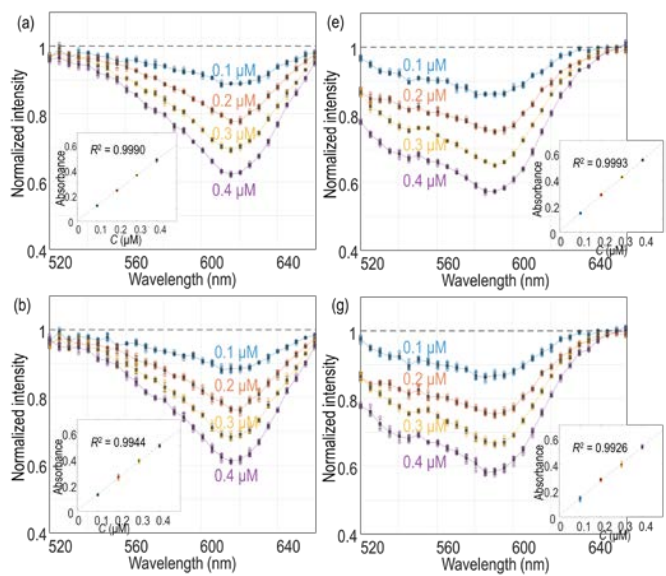

**Fig. 3.** Normalized spectra at the concentrations of 0.1, 0.2, 0.3, and 0.4 μM for (a–d) malachite green and (e–h) crystal violet aqueous solutions. (a), (b), (e), and (f) are the results of measuring five times for the same solution; (c), (d), (g), and (h) are the results of producing five solutions and measuring for each solution. Error bars, standard deviations. (Inset) Absorbance at the dip wavelength as a function of concentrations.

To quantify the enhancement in absorbance (-log($I/I_o$)) of the proposed method, we calculated the ratio of the absorbances between the proposed method and the conventional method (control) using the same sample solution (malachite green and crystal violet, 1 μM) (Figs. 4(a) and 4(b)). For wavelengths near the dips, the enhancement factors were almost constant regardless of the wavelength. The highly variable enhancement in the low-absorbance regimes is due to the near-zero absorbance of the control experiments. The averaged enhancement factors at the wavelengths near the dip were 10.22 times and 10.41 times for malachite green and crystal violet, respectively [Fig. 4(c)]. Because the increment in the OPL is not dependent on the sample, the result of a similar enhancement for the two samples is reasonable.

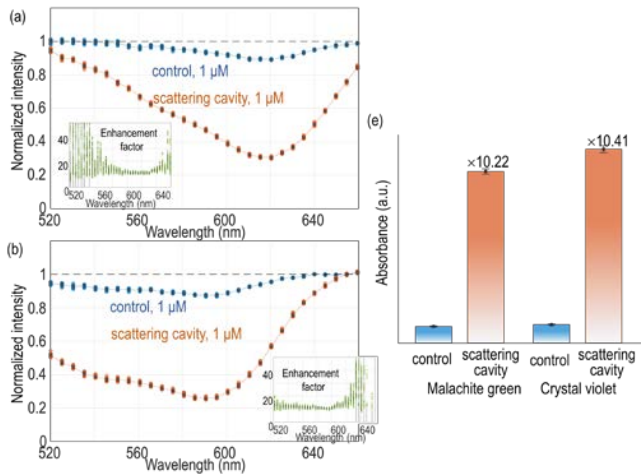

**Fig. 4.** Normalized intensity depending on the wavelength at the concentration of 1 μM for (a) malachite green and (b) crystal violet aqueous solutions using a conventional method and the scattering cavity. At each wavelength, 25 measurements were performed. (Inset) The enhancement factors as a function of the wavelength (c) The averaged enhancement factors near a dip of the normalized intensity for malachite green and crystal violet aqueous solutions. Each plot has error bars with the mean and standard deviation.

To determine the limit of detection (LOD) of the proposed system, we used highly diluted malachite green solutions (Fig. 5). The LOD was obtained by interpolating the concentration that yields the absorbance criterion $\mu_0 + 3\sigma_0$, where $\mu_0$ and $\sigma_0$ are the mean and standard deviation, respectively, of the results using pure water (0 μM) [37, 38].

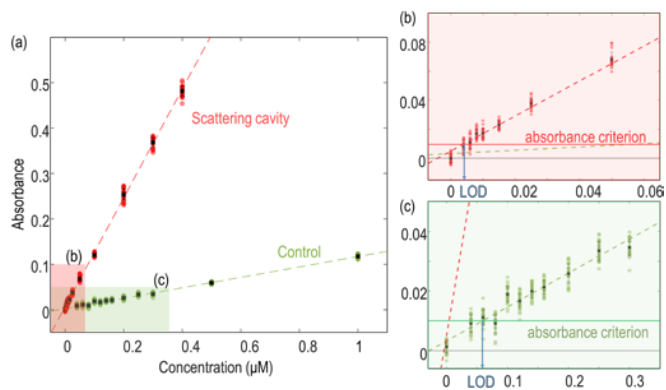

**Fig. 5.** (a) Absorbance in malachite green aqueous solutions of low concentrations measured by the conventional method (green) and the scattering cavity (red). (b-c) Enlarged shaded areas. The measurements were made 25 times each for all concentrations. Each plot has error bars that represent the mean and standard deviation. The absorbance criterion ($\mu_0 + 3\sigma_0$) for estimating the concentration corresponding to the LOD is marked with a blue line.

Figure 5(a) shows the absorbance measurements for the LOD calibration. We employed malachite green solutions at concentrations of 0.040 μM and 0.004 μM for the conventional system (control) and proposed system, respectively. Figures 5(b) and 5(c) show the shaded regions in Fig. 5(a). The absorbance was calculated from the mean and standard deviation of the 0 μM (pure water) absorbance results. We found that the absorbance criteria were $\mu_0 + 3\sigma_0 = 0.01$ in both cases. Note that the absorbance criteria should be the same in both cases because the zero fluctuations are solely derived from the sensitivity and precision of the detection system. In the conventional method (control), the linear correlation was maintained up to 0.120 μM, but the deviation increased for concentrations lower than 0.100 μM (Fig. 5(b)). The calibrated LOD was 0.059 μM, which was consistent with the data observations. In the proposed method, it is noteworthy that the linearity was maintained at substantially lower concentrations (minimum of 0.010 μM) than in the conventional method (Fig. 5(c)). The calibrated LOD was 0.004 μM, which is lower than 1/10 of the LOD of the control.

In summary, we present a scattering cavity to enhance the sensitivity of absorption spectroscopy via a significant increase in the OPL. The scattering cavity was composed of h-BN and led to diffusive reflection on its surface, achieving an improvement in the absorbance by more than 10 times. Moreover, the LOD was drastically lowered in the scattering cavity, and malachite green could be detected even at 0.004 μM using a commercial cuvette. We expect that a higher diffuse reflectance will further improve the enhancement factor. For example, a commercially available highly reflective surface shows >99% diffuse reflectance in the visible and near-infrared range [39]; it can be utilized for applications where a lower LOD is required. The design and size of the scattering cavity can be optimized for further enhancement.

It should be emphasized that the proposed method using multiple scattering is more robust against misalignment than the methods using specular reflection of a mirror. In addition, the proposed method enables quick and simple measurement of absorbance using a plastic cuvette, which is commonly employed in analytical chemistry, and can be readily applicable to existing spectrometers. Due to the effectiveness of the proposed method in detecting trace amounts of a sample mixed in a liquid solvent, the method is expected to widen the detection range for a substance and the wavelength of interest, expanding the research area. We believe that our system can be beneficial for practical applications that require low LOD in water [40], such as those in the food industry [41], pathological diagnosis [42], and biochemical sensing [43, 44].

**Funding.** This work was supported by KAIST UP program, BK21+ program, KAIST Advanced Institute for Science-X, The.Wave.Talk, National Research Foundation of Korea (NRF) (2017M3C1A3013923, 2015R1A3A2066550, 2018K000396, 2018R1A6A3A01011043).

**Disclosures**. The authors declare no competing financial interests.